# Electrically pumped blue laser diodes with nanoporous bottom cladding

Marta Sawicka,[*] Grzegorz Muziol, Natalia Fiuczek, Mateusz Hajdel, Marcin Siekacz, Anna Feduniewicz-Żmuda, Krzesimir Nowakowski-Szkudlarek, Paweł Wolny, Mikołaj Żak, Henryk Turski, and Czesław Skierbiszewski

Institute of High Pressure Physics, Polish Academy of Sciences, Sokołowska 29/37, 01-142 Warsaw, Poland

**ABSTRACT:** We demonstrate electrically pumped III-nitride edge-emitting laser diodes (LDs) with nanoporous bottom cladding. The LD structure was grown by plasma-assisted molecular beam epitaxy. Highly doped 350 nm thick GaN:Si cladding layer with Si concentration of $6 \cdot 10^{19}$ cm$^{-3}$ was electrochemically etched to obtain porosity of 15±3% with pore size of 20±9 nm. The devices with nanoporous bottom cladding are compared to the reference structures. The pulse mode operation was obtained at 448.7 nm with a slope efficiency (SE) of 0.2 W/A while the reference device without etched cladding layer was lasing at 457 nm with SE of 0.56 W/A. The design of the LDs with porous bottom cladding was modelled theoretically. Performed calculations allowed to choose the optimum porosity and thickness of the cladding needed for the desired optical mode confinement and reduced the risk of light leakage to the substrate and to the top-metal contact. This demonstration opens new possibilities for the fabrication of III-nitride LDs.

## INTRODUCTION

Porous GaN emerged recently as an interesting alternative to classical (In,Al,Ga)N alloys in terms or refractive index engineering. Its huge advantage lies in the fact that it provides a way to obtain a reduction of refractive index beyond the possibilities available in nitride material system by increasing material porosity without losing lattice matching. Porous GaN can be obtained in a controllable manner by electrochemical etching (ECE) process using various acidic and neutral electrolytes [1-4]. Obtained material porosity and pore geometry depends on the n-type doping level and applied bias [5, 6]. Two main dependencies can be noted: pore density increases with doping and pore diameter increases with bias [3].

There have been already plenty of applications reported for nanoporous GaN: distributed Bragg reflectors [1, 7], vertical cavity surface emitting lasers [8], strain relief [9], improved light extraction in light-emitting diodes (LEDs) [10], optically pumped edge-emitting lasers [11, 12], photoelectrochemical water-splitting [13, 14], nanomembranes [15, 16], wafer and device lift-off [17, 18] and tracing dopant incorporation during crystal growth [6]. Basic properties of porous GaN have also been studied; the reports include e.g. thermal conductivity [19] and birefringence of a material with aligned pores [20].

The application of nanoporous cladding layers to nitride edge-emitting lasers has been demonstrated in optically pumped structures by Yuan et al. [12]. The authors highlighted the beneficial effect of the decrease of the refractive index in cladding layers and presented more than two-fold reduction of the lasing threshold under optical pumping for the structures with porous claddings. There is also one report on the electrically pumped laser diode

(LD) with a porous bottom cladding grown on a semipolar (20$\bar{2}$1) GaN substrate. However, in this demonstration, the porous layer has been used only for the lateral mode confinement, not for the refractive engineering along the growth direction [21].

In this work we demonstrate blue LDs grown on polar (0001) GaN substrate by plasma-assisted molecular beam epitaxy (PAMBE) that have a porous bottom cladding layer. We disclose the processing scheme and present device structural, electrical and optical characteristics with the comparison to the reference standard LD without etched cladding. Possible routes for the improvement of the performance of LDs with porous cladding are also discussed. In order to illustrate the effect of varied structure refractive index profile on the expected light propagation in the device we calculated the near field intensity profiles, the optical mode confinement factor, $\Gamma$, and light losses to GaN substrate as a function of the bottom cladding porosity and thickness. Impact of other structure parameters such as top cladding thickness on the losses in the upper contact is also shown for LDs with porous bottom claddings. The discussed results point out to the importance of proper LD structure design that provides a high optical mode overlap with the active region.

## MODELLING of the optical mode confinement

In case of nitride material system, the unquestionable advantage - when creating GaN+air "composite" material - comes from the fact that the effective refractive index of porous GaN can be decreased beyond the values available for (In,Al,Ga)N alloys without losing ideal lattice matching to GaN. The refractive index of nanoporous GaN can be naturally tailored down to 1 with increasing porosity. Numerical analysis of the propagation of electromagnetic waves in heterogeneous non-absorbing medium consisting of small air voids showed that the effective index of refraction of a porous material can be successfully modelled using volume averaging theory (VAT).[22] Specifically, this applies to porous GaN with pore diameter, d, less than 30 nm, ensuring the scattering factor $\chi = \pi d/\lambda$ is less than 0.2.[1] Thus, the average refractive index of a porous medium, $n_{por}$, can be expressed as $n_{por} = [(1-\varphi)n_{GaN}^2 + \varphi n_{air}^2]^{1/2}$ where the porosity $\varphi$ is defined as ratio of air volume (pores filled with air) to the total volume [23].

In order to illustrate the extremely high potential of the nanoporous GaN as a cladding layer in blue LDs and find its optimal porosity in our LD structures, we performed calculations of near-field intensity of the fundamental TE mode. We used a model that assumes a free leakage of light out of the waveguide and into the substrate which is a mechanism observed in LDs grown on bulk GaN substrates[24].

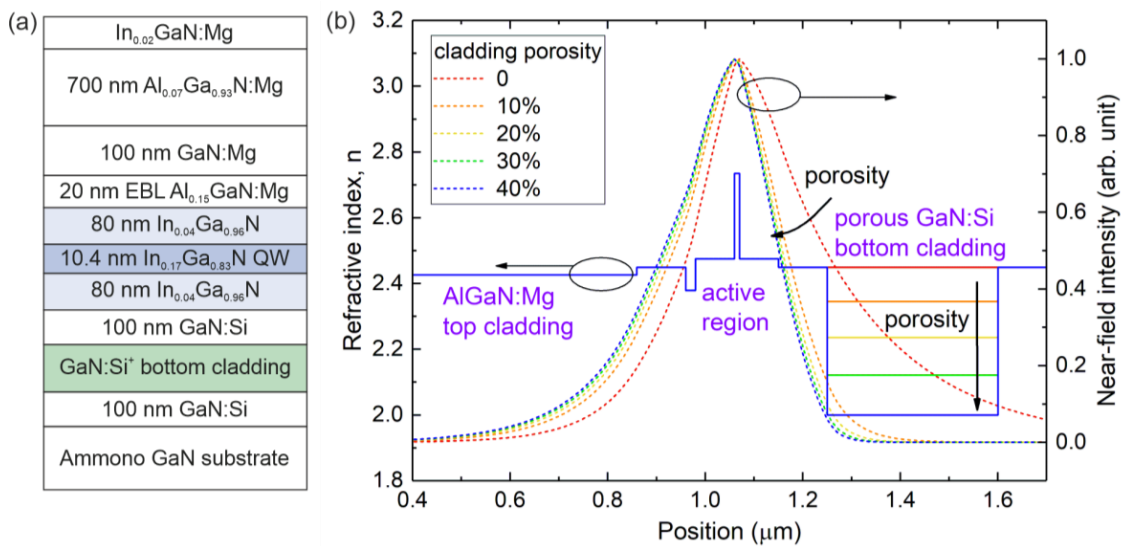

**Figure 1. (a) Layer sequence of the LD structure studied in this work experimentally and theoretically. (b) Refractive index profile and near-field intensity for different LD designs involving 350 nm GaN:Si cladding with porosity from 0 to 40%.**

The layer sequence of the fabricated LDs as well as the structure used for the simulations is presented in Figure 1(a). The structure consists of 100 nm GaN:Si with low doping level $2 \cdot 10^{18}$ cm$^{-3}$ and bottom cladding of 350 nm thick GaN:Si$^+$ that is highly doped to a level of $6 \cdot 10^{19}$ cm$^{-3}$, denoted in green color. Next, there is a 100 nm GaN:Si with low doping level $2 \cdot 10^{18}$ cm$^{-3}$. The porosity and thickness of the highly doped GaN:Si$^+$ is varied in the simulations. The active region of our LD consists of a single quantum well (QW) of 10.4 nm In$_{0.17}$Ga$_{0.83}$N (emission wavelength $\lambda$ = 450 nm), and it is surrounded by a 160 nm In$_{0.04}$Ga$_{0.96}$N waveguide. In the top part, there is also a 20 nm Al$_{0.15}$Ga$_{0.85}$N:Mg electron blocking layer (EBL), 100 nm GaN:Mg layer, 700 nm Al$_{0.07}$Ga$_{0.93}$N:Mg top cladding with doping graded from 1 to $5 \cdot 10^{18}$ cm$^{-3}$ and 45 nm In$_{0.02}$Ga$_{0.98}$N:Mg contact layer. The thickness of the top p-AlGaN cladding in the LD structure was 700 nm while in the simulations it was either 700 or 500 nm.

Figure 1(b) shows the refractive index profile and respective near-field intensity for a reference LD with GaN:Si bottom cladding (denoted with red colour) and LDs with porous bottom cladding with porosity of 10, 20, 30 and 40%, denoted in orange, yellow, green and blue, respectively. The thickness of the bottom cladding is 350 nm. Optical mode, plotted with dashed lines, is positioned at the active region, but the bottom cladding design impacts its symmetry and introduces its shift towards the top-part of the structure when porous cladding layer are used. Comparing the reference LD structure (red dashed line) with the LD with bottom cladding of 10% porosity (orange dashed line), the change is the most significant: the optical mode gets evidently narrower and it is pushed towards the top-part of the structure. Further increase in the porosity only slightly shifts the optical mode.

The detailed change of the influence of the cladding porosity on the confinement factor is shown in Figure 2(a). When the porosity changes from 0 to 10%, $\Gamma$ increases from 2.9 to 3.8. Further increase in the porosity from 20 to 50% causes a slight decrease in $\Gamma$ due to increased asymmetry in the cladding strength and as a consequence, optical mode shift towards the upper part of the LD structure. The maximum of the optical mode is no longer positioned at the QW, which causes the reduction of $\Gamma$. Additionally, it causes the risk of increased losses in gold upper contact and p-type layers as well. The inset shows an exemplary scanning electron microscope (SEM) image of a porous GaN layer with porosity of 35%. This porosity was obtained by ECE at 4V of GaN:Si layer with doping level of $6 \cdot 10^{19}$ cm$^{-3}$.[6]

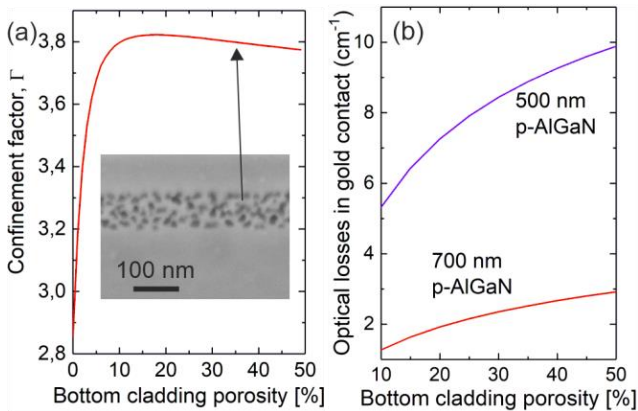

**Figure 2. (a) Confinement factor in LDs as a function of the porosity in the bottom cladding. Exemplary SEM image of a porous GaN layer with porosity 35%. (b) Optical losses in top metal contact for a blue LD with 350 nm GaN porous bottom cladding as a function of its porosity for the LD structure with 700 and 500 nm top p-AlGaN cladding.**

In order to design the upper cladding layer thickness properly, we compared two LD structures with p-AlGaN cladding layers of 500 and 700 nm, keeping the same 350-nm-thick porous GaN cladding layer. Not surprisingly, we find that for the GaN bottom cladding of 10% porosity, we need to provide a sufficiently "strong" confinement on the p-side since the strong asymmetry in the bottom and top cladding (refractive index and thickness) can cause the significant shift of the optical mode towards the upper part of the LD that will result in high light losses in the metal contact and p-type layers. An illustration of this effect is presented in Figure 2(b) where we depict the light losses in the upper gold contact for two cases: 700 and 500 nm p-type Al$_{0.07}$Ga$_{0.93}$N:Mg top cladding as a function of bottom cladding porosity. The higher the porosity in the bottom cladding, the higher the losses in the metal contact. Notably, the losses are larger for a weaker top-cladding (500 nm p-AlGaN) and with increasing porosity they increase more significantly than for the case of stronger top-cladding (700 nm p-AlGaN). Clearly, the 500 nm p-AlGaN is not sufficient to provide the proper light confinement for the studied bottom cladding porosities because the predicted losses are very high, i.e. above 5 cm$^{-1}$.

Cladding layers do not only confine the optical mode around the active region, increasing its overlap with the QW, but also prevent light leakage to GaN substrate. Several approaches to LD designs have been proposed, including application of different bottom cladding materials such as AlGaN, InAlN, their superlattices with GaN, highly n-doped GaN [25-28] or increasing the waveguide refractive index as proposed by Muziol et al.[24]. In this work we analyzed the impact of the bottom cladding design on the light leakage to GaN substrate.

Figure 3 summarizes the predicted light losses in the substrate as a function of porous bottom cladding thickness for the selected porosities: 10, 15, 20 and 25%. We assume that the losses below 2 cm$^{-1}$ can be considered as insignificant. Such low losses can be achieved with 360, 285 , 240 and 208 nm thickness of a porous cladding that has a porosity of 10, 15, 20 or 25%, respectively. A typical thickness of the bottom AlGaN:Si cladding is much larger. In case of violet LD with emission wavelength of $\lambda$=410nm AlGaN cladding as thick as 2 μm has to be used to eliminate the leakage to GaN substrate[29]. In case of longer wavelength LDs, such as emitting at $\lambda$=440nm, the AlGaN thickness has to be increased even up to 3 μm. Note that the thickness of the AlGaN cladding is restricted by the build-up of the compressive strain, eventually causing undesired cracking [30]. Porous cladding, on the other hand, due to its high refractive index contrast to GaN can be much thinner than AlGaN.

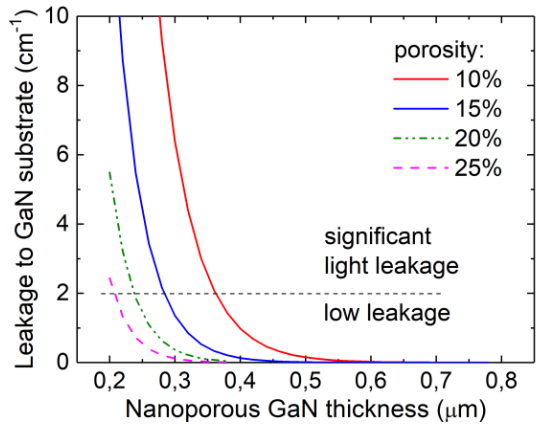

**Figure 3. Light leakage to GaN substrate as a function of the bottom cladding thickness for its different porosity.**

## EXPERIMENTAL

### Epitaxial growth and processing

The epitaxial structure of the LD presented in this work was grown entirely by PAMBE on bulk (0001) GaN. The substrate was an Ammono-GaN crystal with a threading dislocation density ≈5·10$^5$ cm$^{-2}$ [31]. The details of the LD growth conditions have been described in our previous works [32]. The layer sequence is shown in Figure 1(a). Bottom cladding thickness is 350 nm and doping level 6·10$^{19}$ cm$^{-3}$. Importantly, high Si doping needed for the bottom cladding is achieved by PAMBE without problems with deteriorating surface morphology. GaN doping with Si as high as 2·10$^{20}$ cm$^{-3}$ has been reported to be feasible, i.e. providing smooth surface morphology [33, 34]. The capability of achieving high doping of GaN with Si by PAMBE seems to be an advantage over metal-organic vapor phase epitaxy (MOVPE) for which the Si doping higher than 6·10$^{19}$ cm$^{-3}$ is reported to be challenging in terms of preserving high structural quality [35].

After the epitaxy, the LD structure was processed to fabricate ridge-waveguide oxide-isolated lasers. Resonators were 5 μm wide and 700 or 1000 μm long. The device processing was modified and extended by additional steps needed for lateral ECE. A schematic drawing showing the final structure is shown in Figure 4.

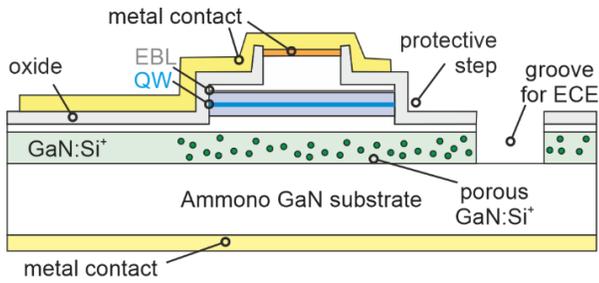

**Figure 4. Schematic of the LD after processing.**

The groove needed for ECE is positioned next to the laser ridge and p-type metallization covers the opposite side of the device. P-type contact metallization consists of two parts, thin and thick, deposited in the first and last processing step, respectively, as previously reported [34, 36]. After the deposition of the thin p-type contact metallization (denoted in orange color in Figure 4), 5-μm-wide laser mesas were formed by reactive ion etching (RIE). Next, additional RIE-step was done to form a protective step for blocking the exposure of the active region and p-type layers to electrolyte. A 250 nm thick silicon oxide ($SiO_2$) was deposited on the surface. Next, 5 μm wide grooves for lateral ECE were etched down through the GaN:Si$^+$ layer by RIE. The grooves were positioned parallel to laser mesa ridge. The distance between groove wall and the middle of the mesa ridge was designed to be 18 μm, therefore the minimum etching distance required to ensure the porous bottom cladding formation beneath the mesa was 20.5 μm. Groove length was intentionally shorter by 100 μm than the device length, i.e. it was 600 (900) μm in case of laser ridge length of 700 (1000) μm. Such design was intended to facilitate mirror cleaving in the area without porous structure. However, due to parasitic channels of etching, it did not work as planned, as it will be shown in the next section.

In order to perform ECE, indium contact was manually placed in the corner of the sample. About 2/3 of the sample area was immersed in the 0.3 M oxalic acid and positively biased at 2.2V for 22.5 hours. Details of the ECE setup is presented in our previous work[6]. Since there is a change in refractive index of the etched layer, it is possible to control the etching front position under optical microscope. The etching distance was 22 μm, that corresponds to the average etching rate of 0.016 μm/min, and seems sufficient to provide complete nanoporous cladding formation beneath the laser ridge. After ECE, the n-side and p-side metallization were deposited. The laser mirrors were cleaved and left uncoated.

## RESULTS

### LD structure characterization

Figure 5(a) presents the top view of an exemplary laser chip. Despite that the whole sample surface was covered by $SiO_2$ (apart from the grooves and their sidewalls), the unintentional etching through the discontinuities in the oxide was locally observed, that we will discuss in the next paragraph. Figure 5(b) presents the SEM image of the porous cladding viewed from the sidewall of the groove. The porosity and pore size of the nanoporous material was estimated using image processing by open source ImageJ application. The porosity is 15 ± 3% and pore size of 20 ± 9 nm. Note that the contrast in SEM image is much better for larger porosity and pore size, as presented in the inset to Figure 2(a). Pore geometry seen on the cleaved surface is shown in Figure 5(c).

LD cross-section in the middle of a device is shown in Figure 5(d). Magnified area of the laser ridge is presented in Figure 5(e). It proves that porous cladding is formed beneath the whole 5-μm-wide laser ridge. The laser mirror roughness is quite high, most likely due to the issues with cleaving related to the presence of porous cladding beneath.

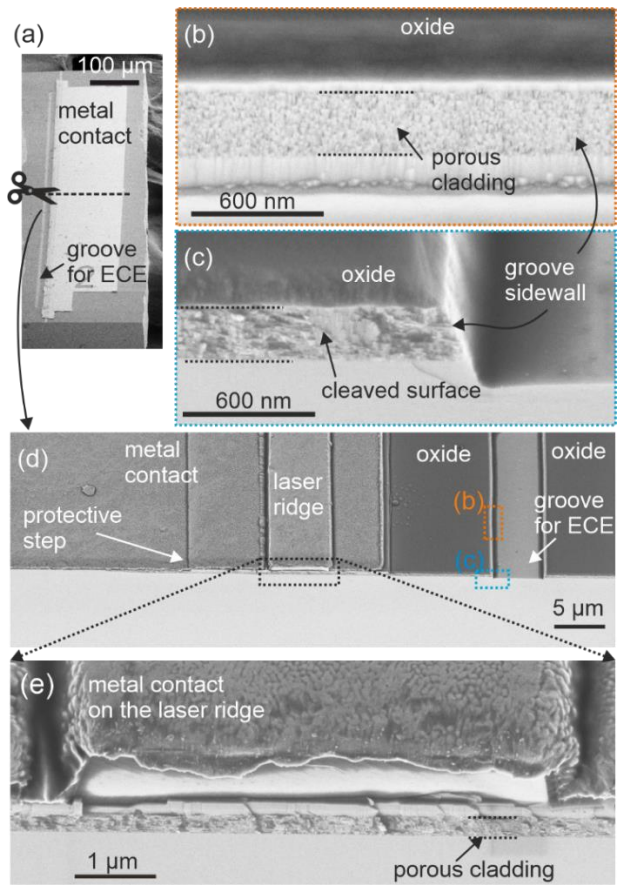

**Figure 5. (a) SEM image of exemplary LD chip. Note the groove length is shorter than the resonator. (b) Porous cladding viewed at the groove sidewall and (c) at the laser cross-section. (d-e) LD cross-section in the middle of a device confirming complete etching of the porous cladding under the laser ridge**.

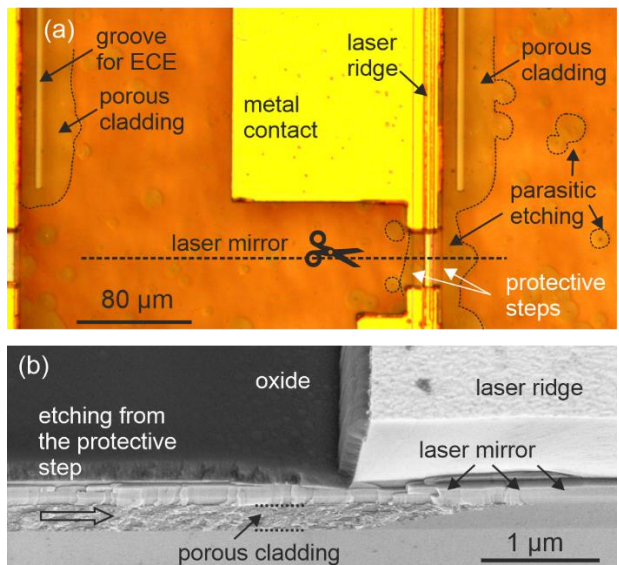

**Figure 6. (a) Optical microscope image of a wafer after processing. Parasitic etching channels around dislocations and locally around the protective steps are marked. (b) Cleaved mirror viewed by SEM at 45° showing that etching proceeded from the protective step towards the laser ridge.**

Figure 6(a) presents the optical microscope image of the processed wafer with a focus on the area between the devices. Despite that the whole sample surface was covered by SiO₂ (apart from the grooves and their sidewalls), the unintentional etching through the discontinuities in the oxide was locally observed. Some of the spots where parasitic etching took place has been indicated in Figure 6(b) by dashed lines. We note two parasitic etching channels: first in the vicinity of the protective step and second related to the presence of dislocations. It was previously shown that dislocation cores may act as effective channels for ECE, allowing mass transport even through the dislocation core diameter as small as 1-2 nm [37].

Additional evidence for parasitic etching through the protective step is presented in Figure 6(b) that shows SEM image of a cleaved mirror of the LD. Cleavage position schematically marked with dashed line in OM image above. The etching front proceeded from the protective step and reaches the laser ridge. Increased mirror roughness can be noted above the porous area.

**Demonstration of LDs with porous cladding**

Room-temperature light-current characteristics for devices with porous and standard bottom claddings are shown in Figure 7(a). They were recorded under pulse mode operation with duty cycle 0.02% and pulse duration of 200 ns for LD with porous cladding and in *cw* mode for the standard LD. Device resonator size was 5×700 μm². The threshold current density for LD with porous cladding was $j_{th}$=9.2 kA/cm² while for the reference structure it was 4.2 kA/cm². Slope efficiency of the LD with porous cladding is 0.2W/A and is lower than for the case of the reference structure 0.56 W/A.

Figure 7(b) presents the emission spectra of the LD with porous cladding recorded under pulse mode and for reference LD with GaN:Si cladding measured in *cw*. The lasing is observed at 448.71 nm for LD with porous cladding and at 456.97 nm for the reference LD without etched cladding, that was cleaved from the same wafer. The difference in laser emission between devices with porous cladding and reference one with GaN:Si cladding is attributed to a partial strain release by the porous cladding causing the blue-shift of the laser emission [9,38].

Figure 7(c) presents the I-V curves of two LDs measured in DC current. The differential resistivity almost identical for both devices, i.e. 6·10⁻⁴ Ω·cm², as denoted by the dashed lines. This is an argument supporting the previous predictions reported by Zhou et al. [19] that the electrical or thermal conductivity of the bottom cladding are not significantly affected by the introduction of 15% porosity. The visible increase of the voltage at a current of 50 mA by about 0.7V can be attributed most likely to the increased p-side contact nonohmicity at low current density.

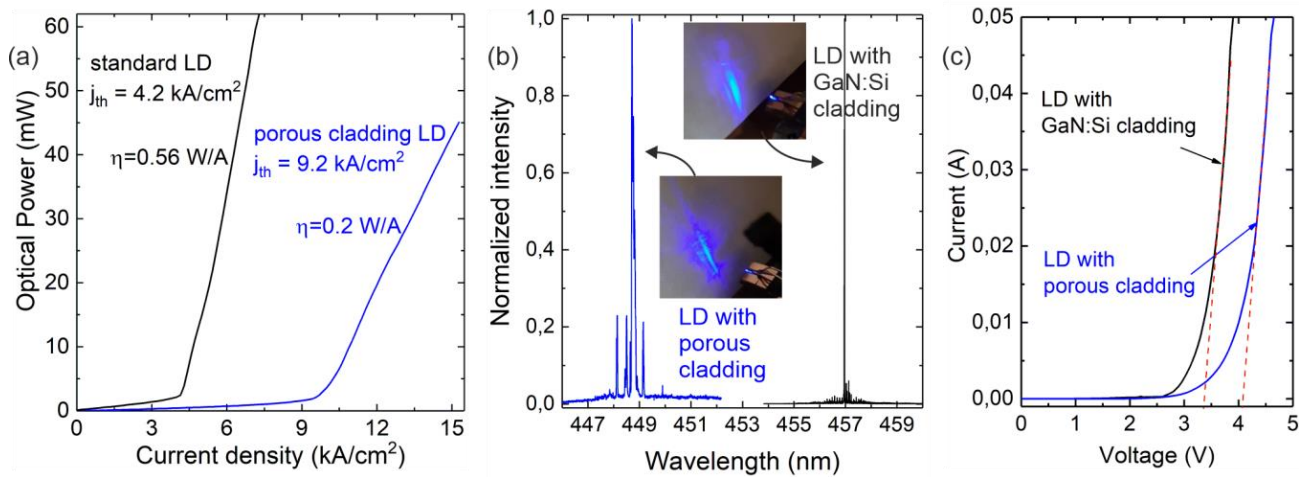

**Figure 7. (a) Light-Current characteristics of the LDs with porous cladding (blue curve) and with standard GaN:Si cladding (black curve). (b) Laser spectra recorded in pulse mode above lasing threshold for LDs with porous cladding λ=448.71 nm and reference LD structure with GaN:Si cladding from the same wafer λ=456.97 nm recorded in cw operation. Insets illustrate the far-field patterns. (c) Current-Voltage characteristics of the LDs with porous cladding (blue curve) and with GaN:Si cladding (black curve). Dashed lines are of the same slope.**

This is likely caused by the probable damage of thin p-type metal contact or InGaN:Mg contact layer during ECE process. Therefore, the need to protect the device structure against parasitic etching is of high importance.

## Discussion

Measured electric and optical properties of the LDs with porous bottom cladding differ quite significantly from their standard counterparts. We expected improvement in the optical mode confinement that should translate into lower threshold current density. However, we observed the opposite. There is about two-fold higher $j_{th}$ and almost a three-fold decrease of slope efficiency for devices with porous cladding. Both of these parameters can be influenced by increased optical losses. However, it is important to note that also injection efficiency can drop at high current densities which are required in order to compensate for the increased optical losses.

Discussing the origins of high threshold current we will analyze the possible sources of the increased losses. We can identify at least five sources of losses: (1) imperfections and increased roughness of the cleaved mirrors, (2) scattering and (3) absorption related to the porous GaN:Si cladding, (4) inhomogeneous mode confinement and (5) absorption in the p-type layers and top metal contact.

Mirror fabrication is definitely one of the most challenging tasks in nitride LD fabrication in general. The losses related to the high mirror surface roughness seem to be inevitable when cleaving is used. Figure 5(e) and Figure 6(b) exemplify cleaved mirror surfaces observed by SEM. The cleavage mirror plane is not perfect, exhibiting large steps. The inability to control the mirror losses of the LDs with porous claddings had definitely a negative impact on the threshold current. This problem can be most likely mitigated if some other mirror fabrication technique is introduced such as focused ion beam (FIB) processing, chemo-mechanical polishing or dry-etching [39, 40]. Additionally, dielectric coatings could be used to further increase mirror reflectivity and improve device performance.

It is very important to stress that there is very little known about possible light scattering on the interface between solid and porous layers. Theoretical works indicate that pore size <30 nm should have a negligible effect on light scattering [22] and the pore size in our structure is estimated to fall below that number. However, more studies are required to shed some light in this regard.

Additionally, the reports on the absorption in the highly n-doped GaN are very limited. There is one report for absorption coefficient measured in a wavelength range from 360 to 400 nm for GaN:Si MOVPE grown layer doped to a low level of $2 \cdot 10^{18}$ cm$^{-3}$ [41] and few studies on oxygen-doped bulk GaN crystals with doping levels from $3 \cdot 10^{19}$ to $1 \cdot 10^{20}$ cm$^{-3}$ grown either by high nitrogen pressure solution or by ammonothermal method [42, 43]. The absorption coefficient is measured as a function of wavelength and doping level. Despite that the absorption coefficient decreases with increasing wavelength in the range from 380 to 450 nm, it still seems high for the strongly n-type doped layers $\approx 6 \cdot 10^{19}$ cm$^{-3}$ that are of our interest. On the other hand, however, the demonstrations of operating LDs with plasmonic claddings emitting at 400 nm [26] support rather the thesis that the losses associated with the absorption in the highly n-doped GaN bottom cladding are probably not very significant.

Third issue that should be taken into account when possible losses are considered is the inhomogeneity in the optical properties of the bottom cladding along the ridge. As mentioned, the groove geometry in respect to the laser ridge was designed in a way that the ends of the ridge would remain solid after ECE and porous structure would be formed across the whole laser ridge apart from $\approx 25$-μm-long ends, cf. Figure 5(a) and Figure 6(a). However, the presence of the parasitic etching channels, mainly at the protective step, caused additional routes for etching, as shown in Figure 6(a-b). We should stress that the ECE of a complete LD structure after epitaxy is challenging due to the severe difficulty in the elimination of parasitic etching channels. Even tiny leaks in the oxide protective layer, e.g. around the dislocations [37], can cause etching of the non-intentionally exposed layers. Therefore, in order to reduce this problem, the use of low dislocation density substrates is a necessity. Additionally, high quality epitaxial growth has to be provided. Another approach that will allow to eliminate the problem of unwanted etching of the QW or p-type involves the epitaxy on a substrate with porous cladding and it will be subject of the future studies.

Last but not least the losses in the p-type layers and upper metal contact could possibly be the issue. According to the simulations, the usage of 700 nm top AlGaN cladding and 350 nm GaN cladding of porosity 15% implies small but nonnegligible optical losses in top metal contact. Also when porous cladding is used, the mode is pushed to the Mg-doped layers which can cause significant losses.[41, 44-46] Both of those aspects could be improved by using not only bottom but also top porous cladding. Porous p-GaN layers have not yet been implemented in any device. The manuscript on the feasibility of porous structure formation in p-type GaN in a controllable manner using tunnel junction as a carrier injection layer is currently under review [47].

We should also note that the possible issues with electrical and thermal conductivity or mechanical stability of porous GaN:Si have proven not to be the limiting factors in the implementation of nanoporous GaN to LD structures. The LDs with porous cladding had satisfactory I-V characteristics as compared to their standard counterparts. Additionally, we did not notice any cracking or structure quality deterioration under optical microscope and SEM inspection despite that some of the processing steps involved an ultrasound bath cleaning.

## CONCLUSIONS

In conclusion, we demonstrated PAMBE-grown electrically pumped III-nitride LDs with porous bottom cladding. By theoretical modelling of the optimum LD structure design we obtained a significant increase in the confinement factor for 350-nm thick GaN:Si bottom cladding with porosity of 15%. Such relatively small porosity provided high mechanical stability and good thermal and electrical conductivity, therefore this design has a potential in practical device applications. In order to obtain LDs with porous claddings via ECE, we implemented additional processing steps in the standard LD processing scheme. The devices with porous cladding were characterized by SEM and their electrical and optical characteristics were compared with the standard structures. The pulse mode operation of porous cladding LDs was obtained at wavelength of 448.7 nm with a slope efficiency (SE) of 0.2 W/A while the reference device without etched cladding layer was lasing at 457 nm with SE of 0.56 W/A. Further work is needed to improve the fabrication process of laser diodes with porous bottom cladding and their properties. This demonstration proves that the concept of porous claddings can be implemented successfully to electrically pumped laser diodes.

## AUTHOR INFORMATION

**Corresponding Author**

* sawicka@unipress.waw.pl.

**Author Contributions**

The manuscript was written through contributions of all authors. All authors have given approval to the final version of the manuscript.

**Funding Sources**

This work received funding from the Foundation for Polish Science co-financed by the European Union under the European Regional Development Fund within the projects POWROTY/REINTEGRATION POIR.04.04.00-00-4463/17-00 and TEAM-TECH POIR.04.04.00-00-210C/16-00. This work was also financially supported by National Science Centre Poland within grants SONATA no. 2019/35/D/ST5/02950 and 2019/35/D/ST3/03008, PRELUDIUM 2019/35/N/ST7/02968 as well as the National Centre for Research and Development within grant no. LIDER/35/0127/L-9/17/NCBR/2018. The research leading to these results has also received funding from the Norway Grants 2014-2021 via the National Centre for Research and Development grant no. NOR/SGS/BANANO/0164/2020.

## ACKNOWLEDGMENT

Authors would like to acknowledge the support of Szymon Stańczyk in high resolution laser spectra acquisition.

## Table of Contents artwork

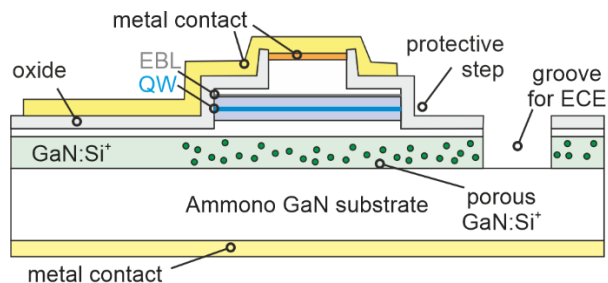